\documentclass[twocolumn,showpacs,preprintnumbers,amsmath,amssymb]{revtex4}
\usepackage{color}
\usepackage{mathrsfs}
\usepackage{amsfonts}
\usepackage{}

\usepackage{graphicx}
\usepackage{dcolumn}
\usepackage{bm}
\usepackage{amsmath,amsthm}
\usepackage{multirow}
\usepackage{array}

\usepackage{amsthm}
\usepackage{graphicx}
\usepackage{dcolumn}

\begin{document}

\title{Impact of the Hawking Effect on the Fully Entangled Fraction of Three-qubit States in Schwarzschild Spacetime}

\author{Guang-Wei Mi$^{1}$}
\author{Xiaofen Huang$^{1}$}
\author{Shao-Ming Fei$^{2}$}
\author{Tinggui Zhang$^{1}$}%
\email{tinggui333@163.com}
\affiliation{
$^{1}$School of Mathematics and Statistics, Hainan Normal University, Haikou 571158, China\\
$^{2}$School of Mathematical Sciences, Capital Normal University, Beijing 100048, China}


\begin{abstract}
Wu \emph{et al.} [J. High Energ. Phys. \textbf{2023}, 232 (2023)] first found that the fidelity of quantum teleportation with a bipartite entangled resource state, completely determined by the fully entangled fraction (FEF) characterized by the maximal fidelity between the given quantum state and the set of maximally entangled states, can monotonically increase in Schwarzschild spacetime. We investigated the Hawking effect on the FEF of quantum states in tripartite systems. In this paper, we show that the Hawking effect of a black hole may both decrease and increase the FEF in Schwarzschild spacetime. For an initial X-type state, we found that the Hawking effect of the black hole has both positive and negative impacts on the FEF of Dirac fields, depending on the selection of initial states. For an initial W-like state, the Hawking effect of the black hole has only a positive impact on the FEF of Dirac fields, independent of the selection of initial states. Our results provide an insightful view of quantum teleportation in multipartite systems under the influence of Hawking effects, from the perspective of quantum information and general relativity.
\end{abstract}

\maketitle


\section{Introduction}
Quantum teleportation plays a vital role in quantum information processing, serving as a fundamental concept in various quantum tasks and contributing significantly to the advancement of quantum technologies such as quantum communication, quantum computing, and quantum networks~\cite{Nielsen.2000,Braunstein.2005,Kimble.2008,Weedbrook.2012,Wilde.2013}. Bennett \emph{et al.} first proposed quantum teleportation in 1993~\cite{Bennett.1993}. The core idea of a quantum teleportation protocol is to utilize the characteristics of quantum entanglement to transmit information to spatially separated receivers while achieving information concealment~\cite{Pirandola.2015}. The basic scheme of quantum teleportation serves as a fundamental component in the advancement of various quantum technologies, including quantum repeaters~\cite{Briegel.1998}, quantum gate teleportation~\cite{Gottesman.1999}, measurement-based quantum computing~\cite{Raussendorf.2001}, and port-based teleportation~\cite{Ishizaka.2008}.

Let $H$ be a $d$-dimensional Hilbert space. If a bipartite quantum state $\rho\in H\otimes H$ is used as a physical resource in quantum teleportation, the fidelity of the teleportation reads $F(\rho)=\frac{f(\rho)d+1}{d+1}$, where $f(\rho)$ is the fully entangled fraction (FEF) \cite{spop}, $f(\rho)=\mathop{\max}_{\phi}\langle\phi|\rho|\phi\rangle$, with the maximization over all maximally entangled states $|\phi\rangle$ \cite{cdjw}.

The development of black-hole physics can be traced to the early 20th century. In 1915, Albert Einstein proposed the theory of general relativity, which describes the geometric nature of gravity. The equations of general relativity predicted that objects with mass concentrated in a very small region would form black holes. Although progress has been made in the field of black-hole physics, many questions remain unresolved, such as the black-hole information paradox~\cite{Danielsson.1993,Anglin.1995,Giddings.2002}, black-hole singularity~\cite{Alwis.1993,Anderson.1993} and event horizon structure~\cite{Jensen.1995,Cvetic.1997}.

In recent decades, the theory of relativistic quantum information has emerged in an attempt to address the problem of unifying general relativity and quantum mechanics~\cite{Asher.2004}. The Hawking effect of a black hole has a negative impact on quantum steering, entanglement, discord, coherence, and the fidelity of quantum teleportation for bosonic fields in the context of curved spacetime~\cite{Fuentes-Schuller.2005,Alsing.2006,Pan.2008,Martn-Martnez.2010,Wang.2010,Esfahani.2011,Xu.2014,Wang.2014,Bruschi.2014,D.E.Bruschi.2014,Hosseinidehaj.2015,
Wang.2016,He.2016,Huang.2018,Shahbazi.2020,Wu.2022,Bhattacharya.2022,Zeng.2022,Wushumin.2022,txsf,J. Ariadna,Qiang,Dong19,Dong20}. For example, Torres-Arenas \emph{et al.} presented the entanglement measures of tripartite W-states in a noninertial frame through the coordinate transformation between Minkowski and Rindler~\cite{J. Ariadna}. Qiang \emph{et al.} presented analytical concurrences for bipartite and tripartite entanglements simultaneously of Dirac fields in noninertial frames~\cite{Qiang}.
However, Wu \emph{et al.} proposed a different viewpoint~\cite{Wu.2023}. They found that as the Hawking temperature increases, the fidelity of quantum teleportation may increase monotonically, rather than necessarily decrease monotonically in a bipartite system. Naturally, we are curious about the impact of the Hawking effect of black holes on teleportation fidelity in tripartite systems. However, in many-body systems, the fidelity of quantum teleportation has not been fully established.

The aim of this study was to generalize the FEF of bipartite systems to the multipartite case. We explored the FEF of multipartite states for Dirac fields in Schwarzschild spacetime. We suppose that Alice, Bob, and Charlie initially share an X-type state or a W-like state. Alice and Bob remain stationary in an asymptotically flat region, while Charlie positions himself near the event horizon of the black hole. For an X-type state, we discovered that the Hawking effect of the black hole has both positive and negative impacts on the FEF of Dirac fields, depending on the selection of the initial states. For a W-like state, the Hawking effect of the black hole has only a positive impact on the FEF of Dirac fields, independent of the selection of the initial states. Consequently, the Hawking effect of the black hole can not only decrease but also increase the FEF in Schwarzschild spacetime.

The rest of this paper is organized as follows. Section \uppercase\expandafter{\romannumeral2} discusses the calculation of the multipartite FEF and the lower bounds of $N$-qubit states. In particular, we introduce the FEF for X-type states of tripartite systems. Section \uppercase\expandafter{\romannumeral3} presents our investigation of the influence of the Hawking effect on the FEF with the X-type state in Schwarzschild spacetime. As discussed in Section \uppercase\expandafter{\romannumeral4}, we investigated the influence of the Hawking effect on the FEF with the W-like state in Schwarzschild spacetime. Conclusions are presented in Section \uppercase\expandafter{\romannumeral5}.

\section{MULTIPARTITE FEF OF $N$-QUBIT STATES}
\newtheorem{definition}{Definition}
\newtheorem{lemma}{Lemma}
\newtheorem{theorem}{Theorem}
\newtheorem{corollary}{Corollary}

\def\QEDclosed{\mbox{\rule[0pt]{1.3ex}{1.3ex}}}
\def\QED{\QEDclosed}
\def\proof{\indent{\em Proof}.}
\def\endproof{\hspace*{\fill}~\QED\par\endtrivlist\unskip}
\newtheorem{example}{\indent Example}

We consider the $N$-qubit $(N\geq2)$ systems $H^{\otimes N}$ with $d=2$. Denote $\{|0\rangle,|1\rangle\}$ the computational basis of $H$. Then, the multipartite FEF of the $N$-qubit state $\rho$ in $H^{\otimes N}$ is given by~\cite{Xujianwei.2016}
\begin{eqnarray}
f(\rho)=\mathop{\max}_{U_{1},\ldots,U_{N}}\langle\phi|(\otimes^{N}_{i=1}U^{\dagger}_{i})\rho(\otimes^{N}_{i=1}U_{i})|\phi\rangle,
\end{eqnarray}
where max runs over all $2\times2$ unitary matrices $U_{1},U_{2},$ $\ldots,U_{N}$ and $|\phi\rangle$ is the GHZ state,
\begin{eqnarray}
|\phi\rangle=\frac{1}{\sqrt{2}}(|00\ldots0\rangle+|11\ldots1\rangle).
\end{eqnarray}
Generally, it is challenging to calculate $f(\rho)$ analytically. We can calculate the lower bounds given by
\begin{eqnarray}
\begin{aligned}
f(\rho)&\geq  \mathop{\max}\limits_{m_{k}}\langle\phi|(\sigma_{m_{1}}\otimes\cdots\otimes\sigma_{m_{N}})\rho(\sigma_{m_{1}}\otimes\cdots\otimes\sigma_{m_{N}})|\phi\rangle\\ & =: f_l(\rho),
\end{aligned}
\end{eqnarray}
where $m_{k}\in\{0,1,2,3\}$, $\sigma_{0}=I$, and $\sigma_{1}=X$, $\sigma_{2}=Y$ and $\sigma_{3}=Z$ are the standard Pauli matrices.

\begin{example}
\label{example-1}
Consider the FEF of the bipartite X-type state,
\begin{eqnarray}
\label{ex1}
\rho_{X}=
\left (
\begin{array}{cccc}
\rho_{11}   &0           &0           &-\rho_{14}\\
0           &\rho_{22}   &-\rho_{23}  &0         \\
0           &-\rho_{23}  &\rho_{33}   &0         \\
-\rho_{14}  &0           &0           &\rho_{44} \\
\end{array}
\right ),
\end{eqnarray}
where $\rho_{11}+\rho_{22}+\rho_{33}+\rho_{44}=1$, $\rho_{22}\rho_{33}\geq|\rho_{23}|^{2}$ and $\rho_{11}\rho_{44}\geq|\rho_{14}|^{2}$.
\end{example}
From (3), we obtain
$f_l(\rho_{X})=\mathop{\max}\langle\phi|(\sigma_{m_{1}}\otimes\sigma_{m_{2}})\rho(\sigma_{m_{1}}\otimes\sigma_{m_{2}})|\phi\rangle$,
where $|\phi\rangle=\frac{1}{\sqrt{2}}(|00\rangle+|11\rangle)$ and max runs over all Pauli matrices $\sigma_{i}$, $i=0,1,2,3$.
We obtain (see Appendix \ref{FEF-ex1})
\begin{eqnarray*}
\begin{aligned}
f_l(\rho_{X})&=\frac{1}{2}(\langle00|+\langle11|)(\sigma_{0}\otimes\sigma_{2})\rho(\sigma_{0}\otimes\sigma_{2})(|00\rangle+|11\rangle)\\
&=\frac{1}{2}(\rho_{22}+\rho_{33}+2\rho_{23}).
\end{aligned}
\end{eqnarray*}
In this case, $m_{1}=0$ and $m_{2}=2$. In fact, the lower bound $f_l$ obtained here is the same as the real $f$ in~\cite{Badziag.2000}.

\begin{example}
\label{example-2}
Consider the following tripartite X-type states,
\begin{eqnarray}
\label{X-type}
\rho_{X}=
\left (
\begin{array}{cccccccc}
\rho_{11}   &0           &0           &0         &0         &0         &0         &-\rho_{18}\\
0           &\rho_{22}   &0           &0         &0         &0         &0         &0         \\
0           &0           &\rho_{33}   &0         &0         &0         &0         &0         \\
0           &0           &0           &\rho_{44} &0         &0         &0         &0         \\
0           &0           &0           &0         &\rho_{55} &0         &0         &0         \\
0           &0           &0           &0         &0         &\rho_{66} &0         &0         \\
0           &0           &0           &0         &0         &0         &\rho_{77} &0         \\
-\rho_{18}  &0           &0           &0         &0         &0         &0         &\rho_{88}
\end{array}
\right ),
\end{eqnarray}
where $\rho_{11}+\rho_{22}+\rho_{33}+\rho_{44}+\rho_{55}+\rho_{66}+\rho_{77}+\rho_{88}=1$ and we assume $\rho_{11}\rho_{88}\geq|\rho_{18}|^{2}$.
\end{example}
From (3), the FEF is estimated by
\begin{eqnarray}\label{7}
f_l(\rho_{X})=\frac{1}{2}(\rho_{11}+\rho_{88}+2\rho_{18}),
\end{eqnarray}
with $\rho_{11}+\rho_{88}\geq\frac{1}{2}$ and $\rho_{18}\geq\frac{1}{2}(1-\rho_{11}-\rho_{88})$; see the detailed analysis in Appendix \ref{FEF-X}.

\section{HAWKING EFFECT ON THE FEF OF X-TYPE STATE IN SCHWARZSCHILD SPACETIME}
We considered the Hawking effect on the FEF of the X-type state in Schwarzschild spacetime. The X-type states are of particular significance in quantum information, especially in the research of quantum entanglement. Rau has made excellent contributions to the study of X-type states and provided the algebraic characterization of X-type states~\cite{Rau.2009}. Moreover, in 2010 Rau studied the generalized $N$-qubit X-type states and their symmetries~\cite{Rau.2010}, giving rise to a comprehensive understanding of the properties of the X-type states.

Assume that Alice, Bob, and Charlie initially share an X-type state for three Unruh modes at an asymptotically flat region of a Schwarzschild black hole; then, Alice and Bob remain in the asymptotically flat region, while Charlie lingers near the event horizon of the black hole. Charlie uses his excited detector to probe the thermal Fermi-Dirac particle distribution. The Unruh vacuum state and the excited state of the fermionic mode in the Schwarzschild spacetime can be written as \cite{Wu.2023}
\begin{eqnarray}
\begin{aligned}
\label{eq15}
|0\rangle_{u}&=\frac{1}{e^{-\frac{\omega}{T}}+1}|0000\rangle-\frac{1}{\sqrt{e^{\frac{\omega}{T}}+e^{-\frac{\omega}{T}}+2}}|0101\rangle\\
&+\frac{1}{\sqrt{e^{\frac{\omega}{T}}+e^{-\frac{\omega}{T}}+2}}|1010\rangle-\frac{1}{e^{\frac{\omega}{T}}+1}|1111\rangle
\end{aligned}
\end{eqnarray}
and
\begin{eqnarray}
\begin{aligned}
\label{eq16}
|1\rangle_{u}&=q_{R}[\frac{1}{\sqrt{e^{-\frac{\omega}{T}}+1}}|1000\rangle-\frac{1}{\sqrt{e^{\frac{\omega}{T}}+1}}|1101\rangle]\\
&+q_{L}[\frac{1}{\sqrt{e^{-\frac{\omega}{T}}+1}}|0001\rangle+\frac{1}{\sqrt{e^{\frac{\omega}{T}}+1}}|1011\rangle],
\end{aligned}
\end{eqnarray}
where $T=\frac{1}{8\pi M}$ is the Hawking temperature with $M$ the mass of the black hole. $\omega$ is frequency. $|mm^{'}n^{'}n\rangle=|m_{k}\rangle^{+}_{out}|m^{'}_{-k}\rangle^{-}_{in}\\|n^{'}_{-k}\rangle^{-}_{out}|n_{k}\rangle^{+}_{in}$, where
$\{|m_{k}\rangle^{+}_{out}\}$ and $\{|m_{-k}\rangle^{-}_{in}\}$ are the orthonormal bases for the exterior and interior regions (denoted by the subscripts $\{out,in\}$) of the Schwarzschild black hole, respectively. The superscripts $\{+,-\}$ represent fermions and anti-fermions, respectively. The coefficients $q_{R}$ and $q_{L}$ in (\ref{eq16}) satisfy $|q_{R}|^{2}+|q_{L}|^{2}=1$.

Because Charlie is unable to access the modes within the event horizon of the black hole, we trace out the inaccessible modes. The reduced density matrix $\rho^{ABC_{out}}_{X}$ is of the type (\ref{X-type}), with all the nonzero entries given by
\begin{eqnarray*}
\begin{aligned}
&\rho^{ABC_{out}}_{1,1}=(e^{-\frac{\omega}{T}}+1)^{-1}\rho_{11}+|q_{L}|^{2}(e^{-\frac{\omega}{T}}+1)^{-1}\rho_{22},\\
&\rho^{ABC_{out}}_{3,3}=|q_{R}|^{2}\rho_{22},\\
&\rho^{ABC_{out}}_{4,4}=(e^{\frac{\omega}{T}}+1)^{-1}\rho_{11}+|q_{L}|^{2}(e^{\frac{\omega}{T}}+1)^{-1}\rho_{22},\\
&\rho^{ABC_{out}}_{5,5}=(e^{-\frac{\omega}{T}}+1)^{-1}\rho_{33}+|q_{L}|^{2}(e^{-\frac{\omega}{T}}+1)^{-1}\rho_{44},\\
&\rho^{ABC_{out}}_{7,7}=|q_{R}|^{2}\rho_{44},\\
&\rho^{ABC_{out}}_{8,8}=(e^{\frac{\omega}{T}}+1)^{-1}\rho_{33}+|q_{L}|^{2}(e^{\frac{\omega}{T}}+1)^{-1}\rho_{44},\\
&\rho^{ABC_{out}}_{9,9}=(e^{-\frac{\omega}{T}}+1)^{-1}\rho_{55}+|q_{L}|^{2}(e^{-\frac{\omega}{T}}+1)^{-1}\rho_{66},\\
&\rho^{ABC_{out}}_{11,11}=|q_{R}|^{2}\rho_{66},\\
&\rho^{ABC_{out}}_{12,12}=(e^{\frac{\omega}{T}}+1)^{-1}\rho_{55}+|q_{L}|^{2}(e^{\frac{\omega}{T}}+1)^{-1}\rho_{66},\\
&\rho^{ABC_{out}}_{13,13}=(e^{-\frac{\omega}{T}}+1)^{-1}\rho_{77}+|q_{L}|^{2}(e^{-\frac{\omega}{T}}+1)^{-1}\rho_{88},\\
&\rho^{ABC_{out}}_{15,15}=|q_{R}|^{2}\rho_{88},\\
&\rho^{ABC_{out}}_{16,16}=(e^{\frac{\omega}{T}}+1)^{-1}\rho_{77}+|q_{L}|^{2}(e^{\frac{\omega}{T}}+1)^{-1}\rho_{88},\\
&\rho^{ABC_{out}}_{1,4}=(e^{\frac{\omega}{T}}+e^{-\frac{\omega}{T}}+2)^{-\frac{1}{2}}[\rho_{11}+|q_{L}|^{2}\rho_{22}],\\
&\rho^{ABC_{out}}_{5,8}=(e^{\frac{\omega}{T}}+e^{-\frac{\omega}{T}}+2)^{-\frac{1}{2}}[\rho_{33}+|q_{L}|^{2}\rho_{44}],\\
&\rho^{ABC_{out}}_{9,12}=(e^{\frac{\omega}{T}}+e^{-\frac{\omega}{T}}+2)^{-\frac{1}{2}}[\rho_{55}+|q_{L}|^{2}\rho_{66}],\\
&\rho^{ABC_{out}}_{13,16}=(e^{\frac{\omega}{T}}+e^{-\frac{\omega}{T}}+2)^{-\frac{1}{2}}[\rho_{77}+|q_{L}|^{2}\rho_{88}],\\
&\rho^{ABC_{out}}_{1,15}=q_{R}(e^{-\frac{\omega}{T}}+1)^{-\frac{1}{2}}\rho_{18},\\
&\rho^{ABC_{out}}_{4,15}=q_{R}(e^{\frac{\omega}{T}}+1)^{-\frac{1}{2}}\rho_{18},\\
&\rho^{ABC_{out}}_{14}=\rho^{ABC_{out}}_{41},~~~ \rho^{ABC_{out}}_{58}=\rho^{ABC_{out}}_{85},\\
&\rho^{ABC_{out}}_{9,12}=\rho^{ABC_{out}}_{12,9},~~~\rho^{ABC_{out}}_{13,16}=\rho^{ABC_{out}}_{16,13},\\
&\rho^{ABC_{out}}_{1,15}=\rho^{ABC_{out}}_{15,1},~~~\rho^{ABC_{out}}_{4,15}=\rho^{ABC_{out}}_{15,14}.
\end{aligned}
\end{eqnarray*}

Assume that Charlie's detector exclusively detects fermionic modes, i.e., the antifermionic modes remain unexcited in a single detector upon fermion detection. Consequently, by tracing out the antifermionic mode $\{|n^{'}_{-k}\rangle^{-}_{out}\}$ beyond the event horizon of the Schwarzschild black hole, we obtain
\begin{eqnarray}
\label{SX-type}
\rho^{S}_{X}=
\left (
\begin{array}{cccccccc}
\rho^{S}_{11}   &0           &0           &0         &0         &0         &0         &-\rho^{S}_{18}\\
0           &\rho^{S}_{22}   &0           &0         &0         &0         &0         &0         \\
0           &0           &\rho^{S}_{33}   &0         &0         &0         &0         &0         \\
0           &0           &0           &\rho^{S}_{44} &0         &0         &0         &0         \\
0           &0           &0           &0         &\rho^{S}_{55} &0         &0         &0         \\
0           &0           &0           &0         &0         &\rho^{S}_{66} &0         &0         \\
0           &0           &0           &0         &0         &0         &\rho^{S}_{77} &0         \\
-\rho^{S}_{18}  &0           &0           &0         &0         &0         &0         &\rho^{S}_{88} \\
\end{array}
\right ),
\end{eqnarray}
where
\begin{eqnarray*}
\begin{aligned}
&\rho^{s}_{11}=(e^{-\frac{\omega}{T}}+1)^{-1}\rho_{11}+|q_{L}|^{2}(e^{-\frac{\omega}{T}}+1)^{-1}\rho_{22},\\
&\rho^{s}_{22}=(e^{\frac{\omega}{T}}+1)^{-1}\rho_{11}+[|q_{R}|^{2}+|q_{L}|^{2}(e^{\frac{\omega}{T}}+1)^{-1}]\rho_{22},\\
&\rho^{s}_{33}=(e^{-\frac{\omega}{T}}+1)^{-1}\rho_{33}+|q_{L}|^{2}(e^{-\frac{\omega}{T}}+1)^{-1}\rho_{44},\\
&\rho^{s}_{44}=(e^{\frac{\omega}{T}}+1)^{-1}\rho_{33}+[|q_{R}|^{2}+|q_{L}|^{2}(e^{\frac{\omega}{T}}+1)^{-1}]\rho_{44},\\
&\rho^{s}_{55}=(e^{-\frac{\omega}{T}}+1)^{-1}\rho_{55}+|q_{L}|^{2}(e^{-\frac{\omega}{T}}+1)^{-1}\rho_{66},\\
&\rho^{s}_{66}=(e^{\frac{\omega}{T}}+1)^{-1}\rho_{55}+[|q_{R}|^{2}+|q_{L}|^{2}(e^{\frac{\omega}{T}}+1)^{-1}]\rho_{66},\\
&\rho^{s}_{77}=(e^{-\frac{\omega}{T}}+1)^{-1}\rho_{77}+|q_{L}|^{2}(e^{-\frac{\omega}{T}}+1)^{-1}\rho_{88},\\
&\rho^{s}_{88}=(e^{\frac{\omega}{T}}+1)^{-1}\rho_{77}+[|q_{R}|^{2}+|q_{L}|^{2}(e^{\frac{\omega}{T}}+1)^{-1}]\rho_{88},\\
&\rho^{s}_{18}=q_{R}(e^{-\frac{\omega}{T}}+1)^{-\frac{1}{2}}\rho_{18}.\\
\end{aligned}
\end{eqnarray*}

Assuming that
\begin{eqnarray*}
\begin{aligned}
(e^{-\frac{\omega}{T}}+&1)^{-1}\rho_{11}+|q_{L}|^{2}(e^{-\frac{\omega}{T}}+1)^{-1}\rho_{22}+(e^{\frac{\omega}{T}}+1)^{-1}\rho_{77}\\
&+[|q_{R}|^{2}+|q_{L}|^{2}(e^{\frac{\omega}{T}}+1)^{-1}]\rho_{88}\geq\frac{1}{2},
\end{aligned}
\end{eqnarray*}
we obtain from (\ref{7}???)
\begin{eqnarray}
\begin{aligned}
f_l(\rho^{S}_{X})&=\frac{1}{2}\{(e^{-\frac{\omega}{T}}+1)^{-1}\rho_{11}+|q_{L}|^{2}(e^{-\frac{\omega}{T}}+1)^{-1}\rho_{22}\\
&+(e^{\frac{\omega}{T}}+1)^{-1}\rho_{77}+2q_{R}(e^{-\frac{\omega}{T}}+1)^{-\frac{1}{2}}\rho_{18}\\
&+[|q_{R}|^{2}+|q_{L}|^{2}(e^{\frac{\omega}{T}}+1)^{-1}]\rho_{88}\}.
\end{aligned}
\end{eqnarray}
The difference of $f(\rho^{S}_{X})$ between Hawking temperatures $T=T_{0}$ and $T=0$ is given by
\begin{eqnarray*}
\begin{aligned}
\Delta_{T}f_l(\rho^{S}_{X}(T_0))&\equiv f_l(\rho^{S}_{X}(T=T_{0}))-f_l(\rho^{S}_{X}(T=0))\\
&=\frac{1}{2}\{(\rho_{77}-\rho_{11}-|q_{L}|^{2}\rho_{22}+|q_{L}|^{2}\rho_{88})\\
&\cdot(e^{\frac{\omega}{T}}+1)^{-1}-2q_{R}\rho_{18}[1-(e^{-\frac{\omega}{T}}+1)^{-\frac{1}{2}}]\}.
\end{aligned}
\end{eqnarray*}
$\Delta f_l(\rho^{S}_{X}(T_{0}))>0$ ($<0$) implies that $\frac{\partial f_l(\rho^{S}_{X})}{\partial T}|_{T_{0}}>0$ ($<0$).

\begin{figure}[h]
\centering
\includegraphics[scale=0.6]{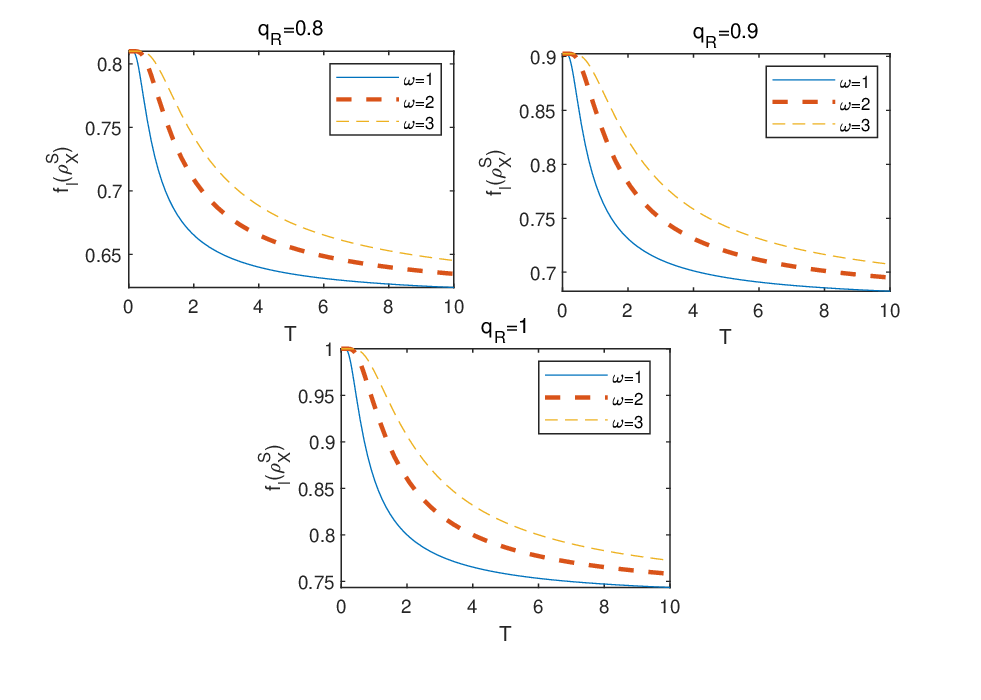}
\caption{Lower bound of FEF $f_l(\rho^{S}_{X})$ as a function of Hawking temperature $T$ for different $\omega$ and $q_{R}$. The initial parameters are fixed as $\rho_{11}=\rho_{88}=\rho_{18}=\frac{1}{2}$ and $\rho_{22}=\rho_{77}=0$, which correspond to the GHZ states.}
\label{Fig1}
\end{figure}

\begin{figure}[h]
\centering
\includegraphics[scale=0.6]{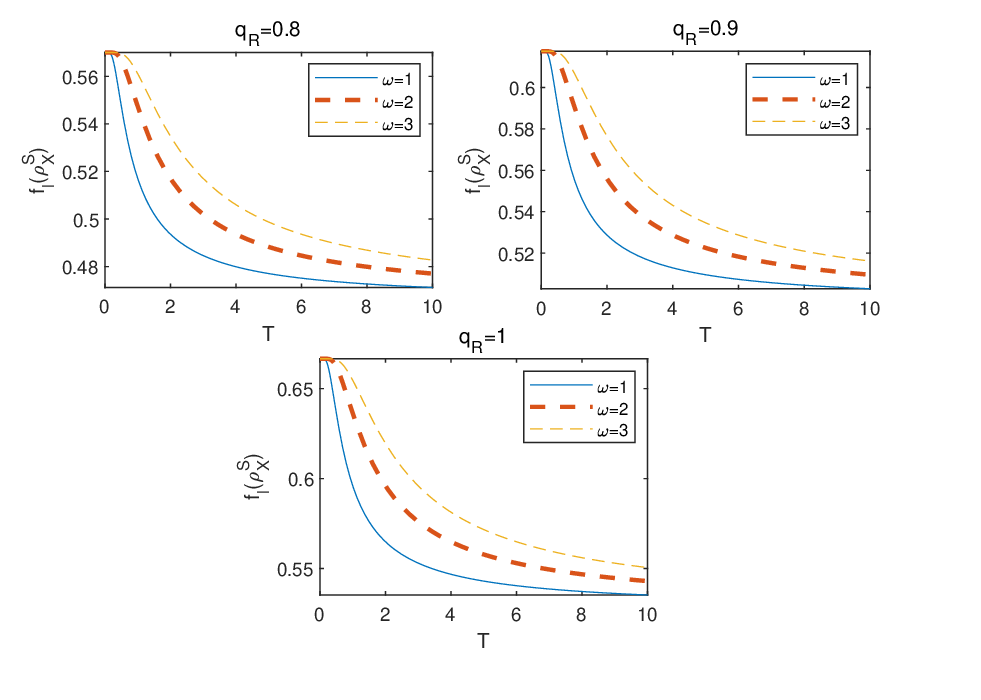}
\caption{Lower bound of FEF $f_l(\rho^{S}_{X})$ as a function of Hawking temperature $T$ for different $\omega$ and $q_{R}$. The initial parameters are fixed as $\rho_{11}=\rho_{88}=\rho_{18}=\frac{1}{3}$ and $\rho_{22}=\rho_{77}=\frac{1}{6}$.}
\label{Fig2}
\end{figure}

Next, we investigated the variation of $f_l(\rho^{S}_{X})$ with respect to the Hawking temperature $T$ through several examples. As shown in Fig.~\ref{Fig1} (GHZ state) and Fig.~\ref{Fig2}, it is evident that the fully entangled fraction $f_l(\rho^{S}_{X})$ exhibits a monotonic decrease with increasing Hawking temperature $T$. Furthermore, we observe that $f_l(\rho^{S}_{X})$ is a monotonically increasing function of the frequency $\omega$ and $q_{R}$. Note that the fully entangled fraction $f_l(\rho^{S}_{X})$ depends on the selection of the Unruh modes. An Unruh mode with $q_{R}=1$ is optimal. Consequently, the FEF can be preserved by selecting the high-frequency mode for maximally entangled states in Schwarzschild spacetime.

\begin{figure}[h]
\centering
\includegraphics[scale=0.6]{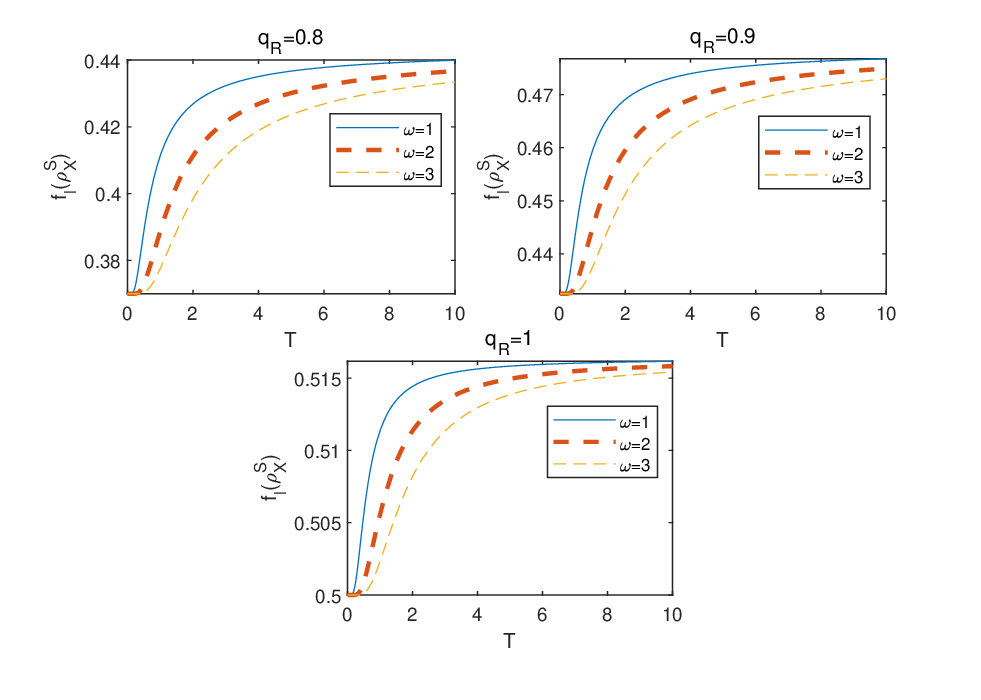}
\caption{Lower bound of FEF $f_l(\rho^{S}_{X})$ as a function of Hawking temperature $T$ for different $\omega$ and $q_{R}$. The initial parameters are fixed as $\rho_{11}=0.1, \rho_{22}=0, \rho_{77}=0.4, \rho_{88}=0.5$ and $\rho_{18}=0.2$.}
\label{Fig3}
\end{figure}

As shown in Fig.~\ref{Fig3}, $f_l(\rho^{S}_{X})$ increases monotonically as the Hawking temperature $T$ increases. This result implies that the Hawking effect of the black hole has a beneficial impact on the FEF. Consequently, the Hawking effect may result in a positive impact on the overall fidelity of quantum teleportation. Additionally, we observe that $f_l(\rho^{S}_{X})$ increases with increasing $q_{R}$. This result again demonstrates that the FEF depends on the choice of the Unruh modes, and the Unruh mode with $q_{R}=1$ is always optimal for the FEF. However, an increase in the frequency $\omega$ has a by-effect on the FEF.

\begin{figure}[h]
\centering
\includegraphics[scale=0.6]{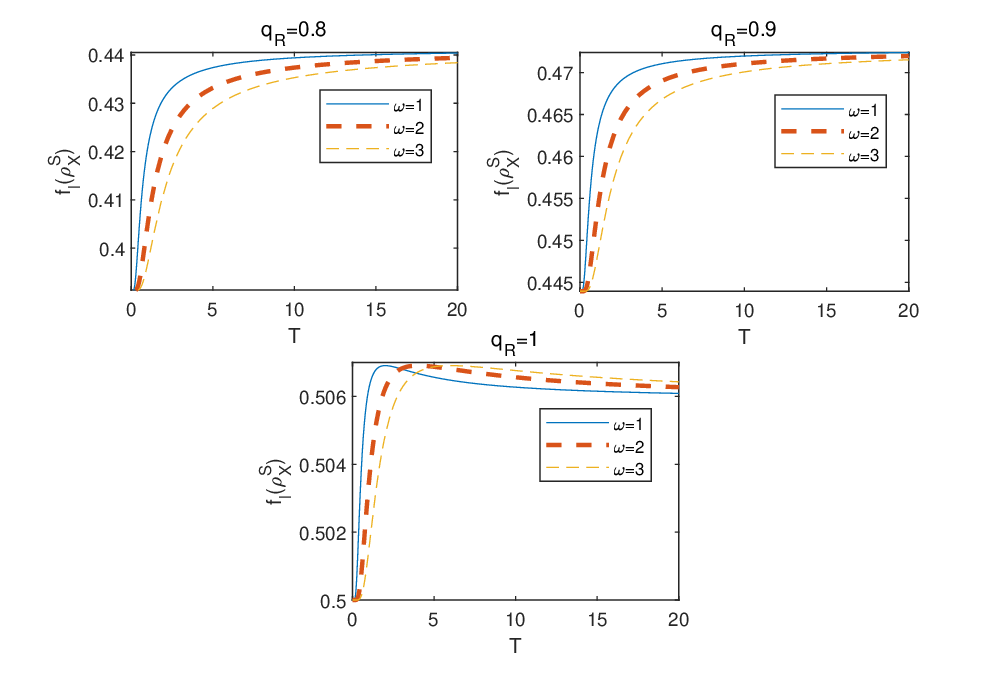}
\caption{Lower bound of FEF $f_l(\rho^{S}_{X})$ as a function of Hawking temperature $T$ for different $\omega$ and $q_{R}$. The initial parameters are fixed as $\rho_{11}=\frac{4-2\sqrt{3}}{3}, \rho_{22}=0, \rho_{77}=\frac{2\sqrt{3}-2}{3}, \rho_{88}=\frac{1}{3}$ and $\rho_{18}=\frac{\sqrt{3}-1}{3}$.}
\label{Fig4}
\end{figure}

As shown in Fig.~\ref{Fig4}, for $q_{R} = 1$, $f_l(\rho^{S}_{X})$ initially rises to its peak value and subsequently decreases consistently as the Hawking temperature $T$ increases. This result indicates that the FEF for the single-mode approximation is both positively and negatively affected by the Hawking effect. It is evident that the maximum fidelity is contingent upon the Hawking temperature $T$ and the frequency $\omega$. Interestingly, for $q_{R} = 0.9$ and $q_{R} = 0.8$, $f_l(\rho^{S}_{X})$ exhibits a monotonic increase as the Hawking temperature $T$ increases. Therefore, we can conclude that for different types of Unruh modes, the FEF exhibits entirely distinct properties with increasing Hawking temperature $T$. That is, the Hawking effect of the black hole has both positive and negative impacts on the FEF of Dirac fields for the X-type state.

\section{HAWKING EFFECT ON THE FEF WITH W-LIKE STATE IN SCHWARZSCHILD SPACETIME}
We considered the following W-like states of the tripartite systems:
\begin{eqnarray*}
\begin{aligned}
\rho_{W}&=\rho_{22}|001\rangle\langle001|+\rho_{23}|001\rangle\langle010|+\rho_{25}|001\rangle\langle100|\\
&+\rho_{32}|010\rangle\langle001|+\rho_{33}|010\rangle\langle010|+\rho_{33}|010\rangle\langle100|\\
&+\rho_{52}|100\rangle\langle001|+\rho_{53}|100\rangle\langle010|+\rho_{55}|100\rangle\langle100|.
\end{aligned}
\end{eqnarray*}
The density matrix is given by
\begin{eqnarray}
\label{W-type}
\rho_{W}=
\left (
\begin{array}{cccccccc}
0           &0           &0           &0         &0         &0         &0         &0\\
0           &\rho_{22}   &\rho_{23}   &0         &\rho_{25} &0         &0         &0\\
0           &\rho_{32}   &\rho_{33}   &0         &\rho_{35} &0         &0         &0\\
0           &0           &0           &0         &0         &0         &0         &0\\
0           &\rho_{52}   &\rho_{53}   &0         &\rho_{55} &0         &0         &0\\
0           &0           &0           &0         &0         &0         &0         &0\\
0           &0           &0           &0         &0         &0         &0         &0\\
0           &0           &0           &0         &0         &0         &0         &0\\
\end{array}
\right ),
\end{eqnarray}
where $\rho_{22}+\rho_{33}+\rho_{55}=1$.

For the state (\ref{W-type}), the lower bound of FEF is given by
\begin{eqnarray}
\begin{aligned}
f_l(\rho_{W})=\frac{1}{2}\rho_{22}
\end{aligned}
\end{eqnarray}
with $\rho_{22}\geq\frac{1}{2}$; see the detailed analysis in Appendix \ref{FEF-W}.

Consider that Alice, Bob, and Charlie initially share a W-like state for three Unruh modes at an asymptotically flat region of the Schwarzschild black hole. Afterwards, Alice and Bob stay in a region that approaches flatness, whereas Charlie remains close to the event horizon of the black hole. Charlie intends to utilize his excited detector to investigate the thermal Fermi-Dirac particle distribution. Because Charlie is unable to access the modes within the event horizon of the black hole, we trace out the inaccessible modes and derive the density matrix $\rho^{ABC_{out}}_{W}$ according to (\ref{eq15}) and (\ref{eq16}); see Appendix \ref{ABCout-W}).

Assume that Charlie's detector exclusively detects fermionic modes, indicating that the antifermionic modes remain unexcited in a single detector upon fermion detection.
Consequently, it is necessary for us to trace out the antifermionic mode $\{|n^{'}_{-k}\rangle^{-}_{out}\}$ beyond the event horizon of the Schwarzschild black hole. Then, we obtain
\begin{eqnarray}
\label{SW-type}
\rho^{S}_{W}=
\left (
\begin{array}{cccccccc}
\rho^{S}_{11} &0               &0              &0         &0         &0             &0         &0\\
0             &\rho^{S}_{22}   &\rho^{S}_{23}  &0         &\rho^{S}_{25} &0         &0         &0\\
0             &\rho^{S}_{32}   &\rho^{S}_{33}  &0         &\rho^{S}_{35} &0         &0         &0\\
0             &0               &0              &\rho^{S}_{44} &0         &\rho^{S}_{46}        &0         &0\\
0             &\rho^{S}_{52}   &\rho^{S}_{53}  &0         &\rho^{S}_{55} &0         &0         &0\\
0             &0               &0              &\rho^{S}_{64}         &0         &\rho^{S}_{66} &0         &0\\
0             &0               &0              &0         &0         &0             &0         &0\\
0             &0               &0              &0         &0         &0             &0         &0
\end{array}
\right ),
\end{eqnarray}
where
\begin{eqnarray*}
\begin{aligned}
&\rho^{s}_{11}=|q_{L}|^{2}(e^{-\frac{\omega}{T}}+1)^{-1}\rho_{22},\\
&\rho^{s}_{22}=[|q_{R}|^{2}+|q_{L}|^{2}(e^{\frac{\omega}{T}}+1)^{-1}]\rho_{22},\\
&\rho^{s}_{33}=(e^{-\frac{\omega}{T}}+1)^{-1}\rho_{33},\\
&\rho^{s}_{44}=(e^{\frac{\omega}{T}}+1)^{-1}\rho_{33},\\
&\rho^{s}_{55}=(e^{-\frac{\omega}{T}}+1)^{-1}\rho_{55},\\
&\rho^{s}_{66}=(e^{\frac{\omega}{T}}+1)^{-1}\rho_{55},\\
&\rho^{s}_{23}=q_{R}(e^{-\frac{\omega}{T}}+1)^{-\frac{1}{2}}\rho_{23},\\
&\rho^{s}_{25}=q_{R}(e^{-\frac{\omega}{T}}+1)^{-\frac{1}{2}}\rho_{25},\\
&\rho^{s}_{32}=q_{R}(e^{-\frac{\omega}{T}}+1)^{-\frac{1}{2}}\rho_{32},\\
&\rho^{s}_{35}=(e^{-\frac{\omega}{T}}+1)^{-1}\rho_{35},\\
&\rho^{s}_{46}=(e^{\frac{\omega}{T}}+1)^{-1}\rho_{35},\\
&\rho^{s}_{52}=q_{R}(e^{-\frac{\omega}{T}}+1)^{-\frac{1}{2}}\rho_{52},\\
&\rho^{s}_{53}=(e^{-\frac{\omega}{T}}+1)^{-1}\rho_{53},\\
&\rho^{s}_{64}=(e^{\frac{\omega}{T}}+1)^{-1}\rho_{53},\\
\end{aligned}
\end{eqnarray*}

Assume that $\rho^{S}_{W}$ satisfies the condition
\begin{eqnarray*}
\begin{aligned}
\rho^{s}_{22}=[|q_{R}|^{2}+|q_{L}|^{2}(e^{\frac{\omega}{T}}+1)^{-1}]\rho_{22}\geq\frac{1}{2}.
\end{aligned}
\end{eqnarray*}
We obtain
\begin{eqnarray}
\begin{aligned}
f_l(\rho^{S}_{W})=\frac{1}{2}[|q_{R}|^{2}+|q_{L}|^{2}(e^{\frac{\omega}{T}}+1)^{-1}]\rho_{22}
\end{aligned}
\end{eqnarray}
and
\begin{eqnarray*}
\begin{aligned}
\Delta_{T}f_l(\rho^{S}_{W}(T))&\equiv f_l(\rho^{S}_{W}(T=T_{0}))-F(\rho^{S}_{W}(T=0))\\
&=\frac{1}{2}|q_{L}|^{2}(e^{\frac{\omega}{T}}+1)^{-1}\rho_{22}.
\end{aligned}
\end{eqnarray*}
Therefore, $\Delta f_l(\rho^{S}_{W}(T_{0}))>0$ implies $\frac{\partial f_l(\rho^{S}_{W})}{\partial T}|_{T_{0}}>0$ and $\frac{\partial f_l(\rho^{S}_{W})}{\partial T}|_{T_{0}}<0$ implies $\Delta f_l(\rho^{S}_{W}(T_{0}))<0$.

\begin{figure}[h]
\centering
\includegraphics[scale=0.6]{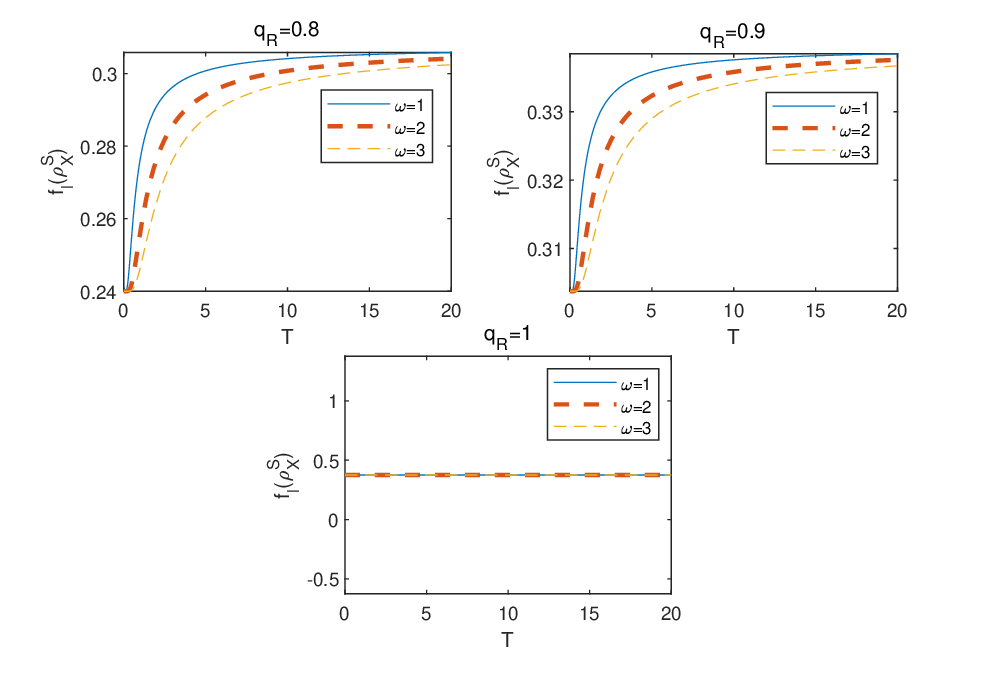}
\caption{Lower bound of FEF $f_l(\rho^{S}_{X})$ as a function of Hawking temperature $T$ for different $\omega$ and $q_{R}$. The initial parameters are fixed as $\rho_{22}=0.75$.}
\label{Fig5}
\end{figure}

We illustrate the fluctuation pattern of FEF $f_l(\rho^{S}_{W})$ with respect to the Hawking temperature $T$ by an example. Fig.~\ref{Fig5} presents the relationship between the lower bound of FEF $f_l(\rho^{S}_{W})$ and the Hawking temperature $T$ for different $\omega$, $q_{R}$ and initial parameters. The result shows that for $q_{R}=1$, $f_l(\rho^{S}_{W})$ is a constant number $\frac{3}{8}$. For $q_{R}=0.9$ and $q_{R}=0.8$, $f_l(\rho^{S}_{W})$ monotonically increases as the Hawking temperature $T$ increases, regardless of the initial parameters. We also discovered that $f_l(\rho^{S}_{W})$ increases with increasing $q_{R}$, and the Unruh mode with $q_{R}=1$ is optimal. However, $f_l(\rho^{S}_{W})$ decreases with increasing $\omega$. Therefore, it is recommended to use low-frequency modes to improve the FEF. We found that the Hawking effect of the black hole has a positive impact on the FEF of Dirac fields for the W-like state.

\section{Conclusion}
In this study, we investigated the lower bound of FEF of Dirac fields among the users in Schwarzschild spacetime for tripartite systems.
Alice, Bob, and Charlie initially share an X-type state or a W-like state. Alice and Bob stay still in an asymptotically flat region, while Charlie situates himself close to the event horizon of the black hole. For an X-type state, we found that the Hawking effect of the black hole has both positive and negative impacts on the FEF of Dirac fields, depending on the selection of the initial states. For a W-like state, the Hawking effect of the black hole has a positive impact on the FEF of Dirac fields, independent of the selection of the initial states. Hence, the Hawking effect of the black hole may both decrease and increase the FEF in Schwarzschild spacetime.

Furthermore, the choice of Unruh modes affects the lower bound of FEF. We found that the Unruh mode with $q_{R}=1$ is always optimal. Furthermore, from Fig.~\ref{Fig4} we can conclude that for different types of Unruh modes, the FEF exhibits entirely distinct properties with increasing Hawking temperature $T$. Wu \emph{et al.}~\cite{Wu.2023} obtained some surprising results that overturned the belief that the Hawking effect of the black hole can only destroy the fidelity of quantum teleportation in a bipartite system. Our results show that the Hawking effect may either decrease or increase the FEF lower bound, which may provide an insightful view from the perspective of quantum information and general relativity, and highlight further studies on the fidelity of many-body quantum states under Hawking effects.

\begin{acknowledgments}
This work was supported by the National Natural Science Foundation of China
(NSFC) under Grant Nos. 12204137, 12075159 and 12171044 and the
specific research fund of the Innovation Platform for Academicians
of Hainan Province under Grant No. YSPTZX202215 and the Hainan
Academician Workstation.
\end{acknowledgments}

\appendix
\section{The FEF of Example \ref{example-1}}
\label{FEF-ex1}
According to (3), we obtained the different lower bounds of FEF for (\ref{ex1}) for different $m_{1}$ and $m_{2}$ as shown in Table \ref{Tabex1}.
\begin{widetext}
\begin{table*}[ht]
\caption{The lower bound of FEF of Example \ref{example-1}.}
\label{Tabex1}
\renewcommand\arraystretch{2}
\centering
\resizebox{\linewidth}{!}{
\begin{tabular}{c|c|c|c|c|c|c|c|c|c|c|c}
\hline
\hline
$m_{1}$        &$m_{2}$     &$FEF$        &$m_{1}$   &$m_{2}$    &$FEF$                 &$m_{1}$    &$m_{2}$      &$FEF$               &$m_{1}$    &$m_{2}$       &$FEF$\\
\hline
$0$         &$0$          &$\frac{1}{2}(\rho_{11}+\rho_{44}-2\rho_{14})$         &$1$        &$0$          &$\frac{1}{2}(\rho_{22}+\rho_{33}-2\rho_{23})$
&$2$        &$0$      &$\frac{1}{2}(\rho_{22}+\rho_{33}+2\rho_{23})$             &$3$        &$0$      &$\frac{1}{2}(\rho_{11}+\rho_{44}+2\rho_{14})$\\
$0$         &$1$          &$\frac{1}{2}(\rho_{22}+\rho_{33}-2\rho_{23})$         &$1$        &$1$          &$-\frac{1}{2}(\rho_{11}+\rho_{44}-2\rho_{14})$
&$2$        &$1$      &$-\frac{1}{2}(\rho_{11}+\rho_{44}+2\rho_{14})$            &$3$        &$1$      &$-\frac{1}{2}(\rho_{22}+\rho_{33}+2\rho_{23})$\\
$0$         &$2$          &$\frac{1}{2}(\rho_{22}+\rho_{33}+2\rho_{23})$         &$1$        &$2$          &$-\frac{1}{2}(\rho_{11}+\rho_{44}+2\rho_{14})$
&$2$        &$2$      &$-\frac{1}{2}(\rho_{11}+\rho_{44}-2\rho_{14})$            &$3$        &$2$      &$-\frac{1}{2}(\rho_{22}+\rho_{33}-2\rho_{23})$\\
$0$         &$3$          &$\frac{1}{2}(\rho_{11}+\rho_{44}+2\rho_{14})$         &$1$        &$3$          &$-\frac{1}{2}(\rho_{22}+\rho_{33}+2\rho_{23})$
&$2$        &$3$      &$-\frac{1}{2}(\rho_{22}+\rho_{33}-2\rho_{23})$            &$3$        &$3$      &$-\frac{1}{2}(\rho_{11}+\rho_{44}-2\rho_{14})$\\
\hline
\hline
\end{tabular}
}
\end{table*}.
\end{widetext}

\section{The FEF of (\ref{X-type})}
\label{FEF-X}
According to (3), we obtained the different lower bounds of FEF for (\ref{X-type}) for different $m_{1}$, $m_{2}$ and $m_{3}$ as shown in Table \ref{Tab1}.
\begin{widetext}
\begin{table*}[ht]
\caption{The lower bounds of FEF of (\ref{X-type}).}
\label{Tab1}
\renewcommand\arraystretch{2}
\centering
\resizebox{\linewidth}{!}{
\begin{tabular}{c|c|c|c|c|c|c|c|c|c|c|c|c|c|c|c}
\hline
\hline
$m_{1}$        &$m_{2}$     &$m_{3}$      &$FEF$     &$m_{1}$       &$m_{2}$     &$m_{3}$      &$FEF$          &$m_{1}$       &$m_{2}$     &$m_{3}$      &$FEF$          &$m_{1}$       &$m_{2}$     &$m_{3}$      &$FEF$\\
\hline
$0$           &$0$         &$0$          &$\frac{1}{2}(\rho_{11}+\rho_{88}-2\rho_{18})$      &$1$        &$0$    &$0$      &$\frac{1}{2}(\rho_{44}+\rho_{55})$
  &$2$    &$0$         &$0$          &$\frac{1}{2}(\rho_{44}+\rho_{55})$                 &$3$      &$0$       &$0$     &$\frac{1}{2}(\rho_{11}+\rho_{88}+2\rho_{18})$\\
$0$           &$0$         &$1$          &$\frac{1}{2}(\rho_{22}+\rho_{77})$                 &$1$        &$0$    &$1$      &$-\frac{1}{2}(\rho_{33}+\rho_{66})$
  &$2$    &$0$         &$1$          &$-\frac{1}{2}(\rho_{33}+\rho_{66})$                &$3$      &$0$       &$1$     &$-\frac{1}{2}(\rho_{22}+\rho_{77})$\\
$0$           &$0$         &$2$          &$\frac{1}{2}(\rho_{22}+\rho_{77})$                 &$1$        &$0$    &$2$      &$-\frac{1}{2}(\rho_{33}+\rho_{66})$
  &$2$    &$0$         &$2$          &$-\frac{1}{2}(\rho_{33}+\rho_{66})$                &$3$      &$0$       &$2$     &$-\frac{1}{2}(\rho_{22}+\rho_{77})$\\
$0$           &$0$         &$3$          &$\frac{1}{2}(\rho_{11}+\rho_{88}+2\rho_{18})$      &$1$        &$0$    &$3$      &$-\frac{1}{2}(\rho_{44}+\rho_{55})$
  &$2$    &$0$         &$3$          &$-\frac{1}{2}(\rho_{44}+\rho_{55})$                &$3$      &$0$       &$3$     &$-\frac{1}{2}(\rho_{11}+\rho_{88}-2\rho_{18})$\\
$0$           &$1$         &$0$          &$\frac{1}{2}(\rho_{33}+\rho_{66})$                 &$1$        &$1$    &$0$      &$-\frac{1}{2}(\rho_{22}+\rho_{77})$
  &$2$    &$1$         &$0$          &$-\frac{1}{2}(\rho_{22}+\rho_{77})$                &$3$      &$1$       &$0$     &$-\frac{1}{2}(\rho_{33}+\rho_{66})$\\
$0$           &$1$         &$1$          &$-\frac{1}{2}(\rho_{44}+\rho_{55})$                &$1$        &$1$    &$1$      &$\frac{1}{2}(\rho_{11}+\rho_{88}-2\rho_{18})$
  &$2$    &$1$         &$1$          &$\frac{1}{2}(\rho_{11}+\rho_{88}+2\rho_{18})$      &$3$      &$1$       &$1$     &$\frac{1}{2}(\rho_{44}+\rho_{55})$\\
$0$           &$1$         &$2$          &$-\frac{1}{2}(\rho_{44}+\rho_{55})$                &$1$        &$1$    &$2$      &$\frac{1}{2}(\rho_{11}+\rho_{88}+2\rho_{18})$
  &$2$    &$1$         &$2$          &$\frac{1}{2}(\rho_{11}+\rho_{88}-2\rho_{18})$      &$3$      &$1$       &$2$     &$\frac{1}{2}(\rho_{44}+\rho_{55})$\\
$0$           &$1$         &$3$          &$-\frac{1}{2}(\rho_{33}+\rho_{66})$                &$1$        &$1$    &$3$      &$\frac{1}{2}(\rho_{22}+\rho_{77})$
  &$2$    &$1$         &$3$          &$\frac{1}{2}(\rho_{22}+\rho_{77})$                 &$3$      &$1$       &$3$     &$\frac{1}{2}(\rho_{33}+\rho_{66})$\\
$0$           &$2$         &$0$          &$\frac{1}{2}(\rho_{33}+\rho_{66})$                 &$1$        &$2$    &$0$      &$-\frac{1}{2}(\rho_{22}+\rho_{77})$
  &$2$    &$2$         &$0$          &$-\frac{1}{2}(\rho_{22}+\rho_{77})$                &$3$      &$2$       &$0$     &$-\frac{1}{2}(\rho_{33}+\rho_{66})$\\
$0$           &$2$         &$1$          &$-\frac{1}{2}(\rho_{44}+\rho_{55})$                &$1$        &$2$    &$1$      &$\frac{1}{2}(\rho_{11}+\rho_{88}+2\rho_{18})$
  &$2$    &$2$         &$1$          &$\frac{1}{2}(\rho_{11}+\rho_{88}-2\rho_{18})$      &$3$      &$2$       &$1$     &$\frac{1}{2}(\rho_{44}+\rho_{55})$\\
$0$           &$2$         &$2$          &$-\frac{1}{2}(\rho_{44}+\rho_{55})$                &$1$        &$2$    &$2$      &$\frac{1}{2}(\rho_{11}+\rho_{88}-2\rho_{18})$
  &$2$    &$2$         &$2$          &$\frac{1}{2}(\rho_{11}+\rho_{88}+2\rho_{18})$      &$3$      &$2$       &$2$     &$\frac{1}{2}(\rho_{44}+\rho_{55})$\\
$0$           &$2$         &$3$          &$-\frac{1}{2}(\rho_{33}+\rho_{66})$                &$1$        &$2$    &$3$      &$\frac{1}{2}(\rho_{22}+\rho_{77})$
  &$2$    &$2$         &$3$          &$\frac{1}{2}(\rho_{22}+\rho_{77})$                 &$3$      &$2$       &$3$     &$\frac{1}{2}(\rho_{33}+\rho_{66})$\\
$0$           &$3$         &$0$          &$\frac{1}{2}(\rho_{11}+\rho_{88}+2\rho_{18})$      &$1$        &$3$    &$0$      &$-\frac{1}{2}(\rho_{44}+\rho_{55})$
  &$2$    &$3$         &$0$          &$-\frac{1}{2}(\rho_{44}+\rho_{55})$                &$3$      &$3$       &$0$     &$-\frac{1}{2}(\rho_{11}+\rho_{88}-2\rho_{18})$\\
$0$           &$3$         &$1$          &$-\frac{1}{2}(\rho_{22}+\rho_{77})$                &$1$        &$3$    &$1$      &$\frac{1}{2}(\rho_{33}+\rho_{66})$
  &$2$    &$3$         &$1$          &$\frac{1}{2}(\rho_{33}+\rho_{66})$                 &$3$      &$3$       &$1$     &$\frac{1}{2}(\rho_{22}+\rho_{77})$\\
$0$           &$3$         &$2$          &$-\frac{1}{2}(\rho_{22}+\rho_{77})$                &$1$        &$3$    &$2$      &$\frac{1}{2}(\rho_{33}+\rho_{66})$
  &$2$    &$3$         &$2$          &$\frac{1}{2}(\rho_{33}+\rho_{66})$                 &$3$      &$3$       &$2$     &$\frac{1}{2}(\rho_{22}+\rho_{77})$\\
$0$           &$3$         &$3$          &$-\frac{1}{2}(\rho_{11}+\rho_{88}-2\rho_{18})$     &$1$        &$3$    &$3$      &$\frac{1}{2}(\rho_{44}+\rho_{55})$
  &$2$    &$3$         &$3$          &$\frac{1}{2}(\rho_{44}+\rho_{55})$                 &$3$      &$3$       &$3$     &$\frac{1}{2}(\rho_{11}+\rho_{88}+2\rho_{18})$\\
\hline
\hline
\end{tabular}
}
\end{table*}.
\end{widetext}

\section{The FEF of (\ref{W-type})}
\label{FEF-W}
According to (3), we obtained the different lower bounds of FEF for $\rho_{W}$ (\ref{W-type}) for different $m_{1}$, $m_{2}$, and $m_{3}$ as shown in Table \ref{Tab2}.
\begin{table*}[ht]
\caption{The lower bounds of FEF for $\rho_{W}$ of (\ref{W-type}).}
\label{Tab2}
\renewcommand\arraystretch{1.5}
\centering
\begin{tabular}{c|c|c|c|c|c|c|c|c|c|c|c|c|c|c|c}
\hline
\hline
$m_{1}$        &$m_{2}$     &$m_{3}$      &$FEF$     &$m_{1}$       &$m_{2}$     &$m_{3}$      &$FEF$          &$m_{1}$       &$m_{2}$     &$m_{3}$      &$FEF$          &$m_{1}$       &$m_{2}$     &$m_{3}$      &$FEF$\\
\hline
$0$           &$0$         &$0$          &$0$                                                &$1$        &$0$    &$0$      &$\frac{1}{2}\rho_{55}$
  &$2$    &$0$         &$0$          &$\frac{1}{2}\rho_{55}$                 &$3$      &$0$       &$0$     &$0$\\
$0$           &$0$         &$1$          &$\frac{1}{2}\rho_{22}$                             &$1$        &$0$    &$1$      &$-\frac{1}{2}\rho_{33}$
  &$2$    &$0$         &$1$          &$-\frac{1}{2}\rho_{33}$                &$3$      &$0$       &$1$     &-$\frac{1}{2}\rho_{22}$\\
$0$           &$0$         &$2$          &$\frac{1}{2}\rho_{22}$                             &$1$        &$0$    &$2$      &$-\frac{1}{2}\rho_{33}$
  &$2$    &$0$         &$2$          &$-\frac{1}{2}\rho_{33}$                &$3$      &$0$       &$2$     &-$\frac{1}{2}\rho_{22}$\\
$0$           &$0$         &$3$          &$0$                                                &$1$        &$0$    &$3$      &$-\frac{1}{2}\rho_{55}$
  &$2$    &$0$         &$3$          &$-\frac{1}{2}\rho_{55}$                &$3$      &$0$       &$3$     &$0$\\
$0$           &$1$         &$0$          &$\frac{1}{2}\rho_{33}$                             &$1$        &$1$    &$0$      &$-\frac{1}{2}\rho_{22}$
  &$2$    &$1$         &$0$          &$-\frac{1}{2}\rho_{22}$                &$3$      &$1$       &$0$     &-$\frac{1}{2}\rho_{33}$\\
$0$           &$1$         &$1$          &$-\frac{1}{2}\rho_{55}$                            &$1$        &$1$    &$1$      &$0$
  &$2$    &$1$         &$1$          &$0$                                    &$3$      &$1$       &$1$     &$\frac{1}{2}\rho_{55}$\\
$0$           &$1$         &$2$          &$-\frac{1}{2}\rho_{55}$                            &$1$        &$1$    &$2$      &$0$
  &$2$    &$1$         &$2$          &$0$                                    &$3$      &$1$       &$2$     &$\frac{1}{2}\rho_{55}$\\
$0$           &$1$         &$3$          &$-\frac{1}{2}\rho_{33}$                            &$1$        &$1$    &$3$      &$\frac{1}{2}\rho_{22}$
  &$2$    &$1$         &$3$          &$\frac{1}{2}\rho_{22}$                 &$3$      &$1$       &$3$     &$\frac{1}{2}\rho_{33}$\\
$0$           &$2$         &$0$          &$\frac{1}{2}\rho_{33}$                             &$1$        &$2$    &$0$      &$-\frac{1}{2}\rho_{22}$
  &$2$    &$2$         &$0$          &$-\frac{1}{2}\rho_{22}$                &$3$      &$2$       &$0$     &-$\frac{1}{2}\rho_{33}$\\
$0$           &$2$         &$1$          &$-\frac{1}{2}\rho_{55}$                            &$1$        &$2$    &$1$      &$0$
  &$2$    &$2$         &$1$          &$0$                                    &$3$      &$2$       &$1$     &$\frac{1}{2}\rho_{55}$\\
$0$           &$2$         &$2$          &$-\frac{1}{2}\rho_{55}$                            &$1$        &$2$    &$2$      &$0$
  &$2$    &$2$         &$2$          &$0$                                    &$3$      &$2$       &$2$     &$\frac{1}{2}\rho_{55}$\\
$0$           &$2$         &$3$          &$-\frac{1}{2}\rho_{33}$                            &$1$        &$2$    &$3$      &$\frac{1}{2}\rho_{22}$
  &$2$    &$2$         &$3$          &$\frac{1}{2}\rho_{22}$                 &$3$      &$2$       &$3$     &$\frac{1}{2}\rho_{33}$\\
$0$           &$3$         &$0$          &$0$                                                &$1$        &$3$    &$0$      &-$\frac{1}{2}\rho_{55}$
  &$2$    &$3$         &$0$          &-$\frac{1}{2}\rho_{55}$                &$3$      &$3$       &$0$     &$0$\\
$0$           &$3$         &$1$          &$-\frac{1}{2}\rho_{22}$                            &$1$        &$3$    &$1$      &$\frac{1}{2}\rho_{33}$
  &$2$    &$3$         &$1$          &$\frac{1}{2}\rho_{33}$                 &$3$      &$3$       &$1$     &$\frac{1}{2}\rho_{22}$\\
$0$           &$3$         &$2$          &$-\frac{1}{2}\rho_{22}$                            &$1$        &$3$    &$2$      &$\frac{1}{2}\rho_{33}$
  &$2$    &$3$         &$2$          &$\frac{1}{2}\rho_{33}$                 &$3$      &$3$       &$2$     &$\frac{1}{2}\rho_{22}$\\
$0$           &$3$         &$3$          &$0$                                                &$1$        &$3$    &$3$      &$\frac{1}{2}\rho_{55}$
  &$2$    &$3$         &$3$          &$\frac{1}{2}\rho_{55}$                 &$3$      &$3$       &$3$     &$0$\\
\hline
\hline
\end{tabular}
\end{table*}.

\section{$\rho^{ABC_{out}}_{W}$}
\label{ABCout-W}
According to (\ref{eq15}) and (\ref{eq16}), we can rewrite (\ref{W-type}). Because Charlie is unable to access the modes within the event horizon of the black hole, we trace over the inaccessible modes and derive the following density matrix $\rho^{ABC_{out}}_{W}$:
\begin{widetext}
\begin{eqnarray*}
\label{SX-type}
\rho^{ABC_{out}}_{W}=
\addtocounter{MaxMatrixCols}{10}
\left (
\begin{array}{cccccccccccccccc}
\rho^{ABC_{out}}_{11}   &0        &0        &\rho^{ABC_{out}}_{14}    &0   &0         &0         &0     &0        &0        &0      &0      &0     &0     &0     &0\\
0           &0          &0        &0        &0      &0       &0       &0   &0         &0         &0     &0        &0        &0      &0      &0\\
0     &0     &\rho^{ABC_{out}}_{33}   &0    &\rho^{ABC_{out}}_{35}    &0   &0         &0   &\rho^{ABC_{out}}_{39} &0  &0 &\rho^{ABC_{out}}_{3,12}  &0     &0     &0    &0 \\
\rho^{ABC_{out}}_{41}   &0        &0        &\rho^{ABC_{out}}_{44}    &0   &0         &0         &0     &0        &0        &0      &0      &0     &0     &0     &0\\
0     &0     &\rho^{ABC_{out}}_{53}   &0    &\rho^{ABC_{out}}_{55}    &0   &0  &\rho^{ABC_{out}}_{58}  &\rho^{ABC_{out}}_{59} &0 &0 &\rho^{ABC_{out}}_{5,12} &0  &0  &0 &0 \\
0           &0          &0        &0        &0      &0       &0       &0   &0         &0         &0     &0        &0        &0      &0      &0\\
0           &0          &0        &0        &0      &0       &0       &0   &0         &0         &0     &0        &0        &0      &0      &0\\
0     &0     &\rho^{ABC_{out}}_{83}   &0    &\rho^{ABC_{out}}_{85}    &0   &0  &\rho^{ABC_{out}}_{88}  &\rho^{ABC_{out}}_{89} &0 &0 &\rho^{ABC_{out}}_{8,12} &0  &0  &0 &0 \\
0     &0     &\rho^{ABC_{out}}_{93}   &0    &\rho^{ABC_{out}}_{95}    &0   &0  &\rho^{ABC_{out}}_{98}  &\rho^{ABC_{out}}_{99} &0 &0 &\rho^{ABC_{out}}_{9,12} &0  &0  &0 &0 \\
0           &0          &0        &0        &0      &0       &0       &0   &0         &0         &0     &0        &0        &0      &0      &0\\
0           &0          &0        &0        &0      &0       &0       &0   &0         &0         &0     &0        &0        &0      &0      &0\\
0     &0    &\rho^{ABC_{out}}_{12,3}   &0   &\rho^{ABC_{out}}_{12,5}  &0   &0  &\rho^{ABC_{out}}_{12,8} &\rho^{ABC_{out}}_{12,9} &0 &0 &\rho^{ABC_{out}}_{12,12} &0 &0 &0 &0 \\
0           &0          &0        &0        &0      &0       &0       &0   &0         &0         &0     &0        &0        &0      &0      &0\\
0           &0          &0        &0        &0      &0       &0       &0   &0         &0         &0     &0        &0        &0      &0      &0\\
0           &0          &0        &0        &0      &0       &0       &0   &0         &0         &0     &0        &0        &0      &0      &0\\
0           &0          &0        &0        &0      &0       &0       &0   &0         &0         &0     &0        &0        &0      &0      &0\\
\end{array}
\right ),
\end{eqnarray*}
\end{widetext}
where
\begin{eqnarray*}
\allowdisplaybreaks[4]
\begin{aligned}
&\rho^{ABC_{out}}_{11}=|q_{L}|^{2}(e^{-\frac{\omega}{T}}+1)^{-1}\rho_{22},\\
&\rho^{ABC_{out}}_{14}=\rho^{ABC_{out}}_{41}=|q_{L}|^{2}(e^{\frac{\omega}{T}}+e^{-\frac{\omega}{T}}+2)^{-\frac{1}{2}}\rho_{22},\\
&\rho^{ABC_{out}}_{33}=|q_{R}|^{2}\rho_{22},\\
&\rho^{ABC_{out}}_{35}=q_{R}(e^{-\frac{\omega}{T}}+1)^{-\frac{1}{2}}\rho_{23},\\
\end{aligned}
\end{eqnarray*}
\begin{eqnarray*}
\begin{aligned}
&\rho^{ABC_{out}}_{38}=q_{R}(e^{\frac{\omega}{T}}+1)^{-\frac{1}{2}}\rho_{23},\\
&\rho^{ABC_{out}}_{39}=q_{R}(e^{-\frac{\omega}{T}}+1)^{-\frac{1}{2}}\rho_{25},\\
&\rho^{ABC_{out}}_{3,12}=q_{R}(e^{\frac{\omega}{T}}+1)^{-\frac{1}{2}}\rho_{25},\\
&\rho^{ABC_{out}}_{44}=|q_{L}|^{2}(e^{\frac{\omega}{T}}+1)^{-1}\rho_{22},\\
&\rho^{ABC_{out}}_{53}=q_{R}(e^{-\frac{\omega}{T}}+1)^{-\frac{1}{2}}\rho_{32},\\
\end{aligned}
\end{eqnarray*}
\begin{eqnarray*}
\begin{aligned}
&\rho^{ABC_{out}}_{55}=(e^{-\frac{\omega}{T}}+1)^{-1}\rho_{33},\\
&\rho^{ABC_{out}}_{58}=\rho^{ABC_{out}}_{85}=(e^{\frac{\omega}{T}}+e^{-\frac{\omega}{T}}+2)^{-\frac{1}{2}}\rho_{33},\\
&\rho^{ABC_{out}}_{59}=(e^{-\frac{\omega}{T}}+1)^{-1}\rho_{35},\\
&\rho^{ABC_{out}}_{5,12}=(e^{\frac{\omega}{T}}+e^{-\frac{\omega}{T}}+2)^{-\frac{1}{2}}\rho_{35},\\
&\rho^{ABC_{out}}_{83}=q_{R}(e^{\frac{\omega}{T}}+1)^{-\frac{1}{2}}\rho_{32},\\
&\rho^{ABC_{out}}_{88}=(e^{\frac{\omega}{T}}+1)^{-1}\rho_{33},\\
&\rho^{ABC_{out}}_{89}=(e^{\frac{\omega}{T}}+e^{-\frac{\omega}{T}}+2)^{-\frac{1}{2}}\rho_{35},\\
&\rho^{ABC_{out}}_{8,12}=(e^{\frac{\omega}{T}}+1)^{-1}\rho_{35},\\
&\rho^{ABC_{out}}_{93}=q_{R}(e^{-\frac{\omega}{T}}+1)^{-\frac{1}{2}}\rho_{52},\\
&\rho^{ABC_{out}}_{95}=(e^{-\frac{\omega}{T}}+1)^{-1}\rho_{53},\\
&\rho^{ABC_{out}}_{98}=(e^{\frac{\omega}{T}}+e^{-\frac{\omega}{T}}+2)^{-\frac{1}{2}}\rho_{53},\\
&\rho^{ABC_{out}}_{99}=(e^{-\frac{\omega}{T}}+1)^{-1}\rho_{55},\\
&\rho^{ABC_{out}}_{9,12}=\rho^{ABC_{out}}_{12,9}=(e^{\frac{\omega}{T}}+e^{-\frac{\omega}{T}}+2)^{-\frac{1}{2}}\rho_{55},\\
&\rho^{ABC_{out}}_{12,3}=q_{R}(e^{\frac{\omega}{T}}+1)^{-\frac{1}{2}}\rho_{52},\\
&\rho^{ABC_{out}}_{12,5}=(e^{\frac{\omega}{T}}+e^{-\frac{\omega}{T}}+2)^{-\frac{1}{2}}\rho_{53},\\
&\rho^{ABC_{out}}_{12,8}=(e^{\frac{\omega}{T}}+1)^{-1}\rho_{53},\\
&\rho^{ABC_{out}}_{12,12}=(e^{\frac{\omega}{T}}+1)^{-1}\rho_{55}.\\
\end{aligned}
\end{eqnarray*}


\nocite{*}

\bibliography{apssamp}

\end{document}